\documentclass[superscriptaddress,nofootinbib, amsmath,amssymb,preprintnumbers,prdfloatfix,twocolumn,PRL]{revtex4-2}
\usepackage{CJK}
\usepackage{graphicx}
\usepackage{dcolumn}
\usepackage{upgreek}
\usepackage{amsmath}
\usepackage{bm}
\usepackage[colorlinks=true,pdfstartview=FitV,breaklinks=true]{hyperref}
\usepackage[dvipsnames,table]{xcolor}
\hypersetup{urlcolor=BlueViolet,
	    citecolor=Plum,
	    linkcolor=PineGreen}
\usepackage{lipsum}	    
\usepackage{titlesec}
\usepackage{etoolbox}
\usepackage[normalem]{ulem}
\usepackage{tabularx}
\usepackage{amssymb}
\usepackage{float}	
\usepackage{caption}
\usepackage{subcaption}
\usepackage{comment}
\usepackage{appendix}
\usepackage{aasmacros}
\usepackage{subcaption}
\usepackage{afterpage}

\definecolor{offblue}{RGB}{23,80,153}

\newcommand{\Ms}{M_\odot}

\newcommand{\htwo}{\mathrm{H_2}}
\newcommand{\Tvir}{T_{\rm vir}}
\newcommand{\Tcmb}{T_{\rm cmb}}

\newcommand{\fbh}{f_\mathrm{BH}}
\newcommand{\fint}{f_\mathrm{BH,in.}}
\newcommand{\fext}{f_\mathrm{BH,ex.}}

\newcommand{\equaref}[1]{Eq.~(\ref{#1})}
\newcommand{\equasref}[2]{Eqs.~(\ref{#1})~and~(\ref{#2})}
\newcommand{\equassref}[3]{Eqs.~(\ref{#1}), (\ref{#2})~and~(\ref{#3})}

\newcommand{\figref}[1]{Fig.~\ref{#1}}

\newcommand{\refref}[1]{Ref.~\cite{#1}}
\newcommand{\refsref}[2]{Refs.~\cite{#1}~and~\cite{#2}}

\begin{document}

\preprint{IPMU23-0052}

\title{Feeding plankton to whales: high-redshift supermassive black holes from tiny black hole explosions}


 \author{Yifan Lu}
 \email{yifanlu@g.ucla.edu}

 \author{Zachary S. C. Picker}
 \email{zpicker@physics.ucla.edu}

\affiliation{Department of Physics and Astronomy, University of California Los Angeles,\\ Los Angeles, California, 90095-1547, USA}

\author{Alexander Kusenko}
\email{kusenko@ucla.edu}

\affiliation{Department of Physics and Astronomy, University of California Los Angeles,\\ Los Angeles, California, 90095-1547, USA}
\affiliation{Kavli Institute for the Physics and Mathematics of the Universe (WPI), The University of Tokyo Institutes for Advanced Study, The University of Tokyo, Chiba 277-8583, Japan}

\begin{abstract}
\noindent 
Recent observations of the high-redshift universe have uncovered a significant number of active galactic nuclei, implying that supermassive black holes (SMBHs) would have to have been formed at much earlier times than expected. Direct collapse of metal-free gas clouds to SMBHs after recombination could help explain the early formation of SMBHs, but this scenario is stymied by the fragmentation of the clouds due to efficient molecular hydrogen cooling.   We show that a subdominant population of tiny, evaporating primordial black holes, with significant clustering in some gas  clouds, can heat the gas sufficiently so that molecular hydrogen is not formed, and direct collapse to to black holes is possible even at high redshifts.
\end{abstract}

\maketitle

\section{Introduction}
\noindent The origin of supermassive black holes (SMBHs) residing in quasars and active galactic nuclei (AGN) at high redshifts has long been an intriguing mystery in astronomy and cosmology~\cite{Inayoshi:2019fun, mayer2018route, volonteri2010formation, wiklind2012first,Bogdan:2023ilu}. In particular, recent observations by the James Webb Space Telescope (JWST)~\cite{JWST2006SSRv..123..485G} of high-redshift ($z\gtrsim 6$) active galactic nuclei~\cite{jwst2023A&A...677A.145U,jwst2023ApJ...953L..29L,jwst2023ApJ...959...39H,jwst2023Natur.619..716C,jwst2023ApJ...942L..17O,jwst2023ApJ...954L...4K,jwst2023ARA&A..61..373F,jwstMaiolino:2023zdu} challenge conventional understandings of the formation mechanisms of these massive black holes, which would need to have formed at even higher redshifts ($z\gtrsim 15$)~\cite{Inayoshi:2019fun,sethi2010supermassive}.

There are a large number of proposed solution to this early SMBH problem, but three common categories of solutions include: i), the `light seeds' scenario, wherein SMBH seeds are produced by Pop~III stars after their death~\cite{Banik:2016qww,2009ApJ...706.1184O, kroupa2020very}; ii), the `direct collapse' scenario, wherein gas clouds directly collapse to supermassive stars or quasi-stars which could seed SMBHs~\cite{Inayoshi:2019fun}, and iii), the primordial black hole (PBH) scenario~\cite{pbh,Hawking:1971ei,Carr:1974nx,Chapline:1975ojl,Carr:2020gox}, wherein black holes which form in the very early universe provide the seeds of SMBH formation~\cite{Bean:2002kx,Duechting:2004dk,Khlopov:2004sc,Dokuchaev:2004kr,Kawasaki:2012kn,Carr:2023tpt}.

Due to the rather small initial mass of the Pop~III remnants, a high accretion rate is needed in the light seeds scenario -- one often needs to invoke super-Eddington accretion in order to explain the mass growth from $\sim 10^2 \Ms$ to $\sim 10^9 \Ms$~\cite{Inayoshi:2019fun}. In the direct collapse scenario, meanwhile, the main challenge is to prevent fragmentation of the gas clouds, since this fragmentation suppresses the accretion of gas onto the central nucleus. This fragmentation is sensitive to the stability of the self-gravitating system under local perturbations and the relation between the local orbital time and the cooling timescale. The first condition is studied in \refref{Toomre:1964zx} in the case of a thin protogalactic disk, and the gravitational stability is captured by the so called Toomre parameter~\cite{mayer2018route, Oh:2001ex}. 

The second condition---sufficient suppression of the gas cooling rate---is more difficult to satisfy. For a cloud of metal-free primordial gas, the formation of molecular hydrogen ($\htwo$) significantly increases the cooling rate. This is because inelastic collisions of $\htwo$ efficiently dissipate energy via rotational and vibrational modes~\cite{lepp1983kinetic, hollenbach1979molecule}. Indeed, numerical simulations confirm that fragmentation is suppressed when there is only an insubstantial amount of $\htwo$ formation~\cite{Bromm:2002hb, Choi:2013kia, 2010MNRAS.402.1249S}. In other words, the formation of $\htwo$ typically implies the failure of the direct collapse to a black hole. \refref{Oh:2001ex, Omukai:2008wv} discuss the formation of $\htwo$ and also the cooling and fragmentation of haloes at the virialization temperature $\Tvir \sim 10^4$ K.

One way to suppress $\htwo$ formation is to introduce either a Lyman-Werner (LW) background with photon energy in the range of $11.2\ \rm eV - 13.6\ \rm eV$, or a near-infrared (NIR) background with photon energy greater than $0.76$ eV in order to directly dissociate the molecular hydrogen~\cite{2010MNRAS.402.1249S,Haiman:1996rc}. Such a background can be realized by a halo with a nearby star-forming galaxy~\cite{dijkstra2008MNRAS.391.1961D,Inayoshi:2019fun}, by dark matter annihilation or  decay~\cite{Biermann:2006bu,Stasielak:2006br,Spolyar:2007qv, Natarajan:2008db, Araya:2013dwa,Friedlander:2022ovf}, and by radiation from superconducting cosmic string loops \cite{Cyr:2022urs}. However, SMBHs which form at very high redshifts may challenge the scenario where this radiation is sourced by star formation. 
Another particularly interesting way to suppress $\htwo$ formation is by introducing additional heating sources, such as heating through a primordial magnetic field \cite{sethi2010supermassive}. 

In this paper we offer an alternative suggestion, that the heating source could be the evaporation of a subdominant population of exploding PBHs embedded within the cloud. Black holes are understood to radiate a nearly blackbody spectrum of fundamental particles in a process known as Hawking radiation~\cite{Hawking:1974rv,Hawking:1974sw}. Evaporation from Hawking radiation means that black holes smaller than the `critical mass', $m\sim 8\times10^{14}~g$, have lifetimes shorter than the age of the universe~\cite{MacGibbon:2007yq,arbey_blackhawk_2019}. In particular, black holes of masses around $M\sim 10^{14}~g$ would be completely evaporating (or, e\textit{xploding}) after recombination and before $z\sim 10$, precisely when we require the additional energy injection for direct collapse. The strongest constraints on the PBH abundance in this mass range comes from the effect of Hawking radiation on the 21-cm line~\cite{Cang:2021owu}, along with other early universe constraints such as cosmic microwave background (CMB) anisotropies and big bang nucleosynthesis (BBN)~\cite{carr_new_2010,Acharya:2020jbv,Chluba:2020oip,Carr:2020gox}.

Clustering of PBH depends on the production scenarios~\cite{Khlopov:2004sc,Clesse:2016vqa,Young:2019gfc,Korol:2019jud,DeLuca:2020jug,DeLuca:2021hcf,Gorton:2022fyb,DeLuca:2022uvz}. Notably, many realistic scenarios of PBH formation feature additional clustering beyond the basic Poisson fluctuations,~\cite{Ferrante:2022mui,Franciolini:2023wun,Young:2019gfc,Carr:2023tpt,Flores:2020drq,Flores:2021jas,Cotner:2016cvr}, so that areas of high density such as the dark matter halos associated to some of these gas clouds might have locally larger PBH fraction. In a number of newly proposed scenarios for PBH formation~\cite{Cotner:2016cvr,Cotner:2017tir,Cotner:2018vug,Cotner:2019ykd,Flores:2020drq,Flores:2021jas}, clustering has not been analyzed in detail, but one can expect it to be significant, especially when the PBH formation involves relatively long-range scalar interactions~\cite{Flores:2020drq,Flores:2021jas}. 
We show that in scenario with large clustering, the exploding PBHs can provide enough heating to prevent the formation of molecular hydrogen and ultimately cause the cloud to collapse directly into a supermassive black hole.

The outline of the paper is as follows. In Sec.~\ref{sec:halo}, we introduce the collapse model of the halo, particularly outlining the chemical and temperature evolution of the cloud. In Sec.~\ref{sec:pbh} we outline the heating of the cloud from the explosion of small primordial black holes. We present the numerical results from the collapse in Sec.~\ref{sec:results} before concluding. We use the Planck~\cite{Planck:2018vyg} values for cosmological constants throughout this work---$H_0 = 68 \ {\rm km/s/Mpc}$, $\Omega_{\Lambda} = 0.69$, $\Omega_{dm} = 0.261$ and $\Omega_{b} = 0.049$.

\section{Halo Evolution}\label{sec:halo}

\noindent We must track the halo temperature and the abundance of the various chemical components during the direct collapse process to determine whether fragmentation of the cloud occurs. In this section, we introduce the analytic model for the collapsing halo and the differential equations for the chemical and temperature evolution used in our numerical code. We assume here that the gas is heated by a component proportional to the dark matter density, and derive the equations required to simulate the numerical collapse in general before introducing the specific PBH-scenario heating rate in Sec.~\ref{sec:pbh}.

\subsection{Initial collapse model}
\noindent For the majority of the collapse process the density perturbation is well beyond the linear regime. The spherical top-hat collapse model \cite{1993sfu..book.....P} offers an analytic model for the density evolution of the cloud in this non-linear regime and has been widely used to study collapse processes before virialization~\cite{Tegmark:1996yt, Omukai:2000ic, Stasielak:2006br, Sethi:2008eq}. In this scheme, the overdensity,
\begin{equation}
    \delta = \frac{\rho^\prime}{\rho} - 1~,
\end{equation}
evolves according to the parametric equation,
\begin{equation}
    1 + \delta = \frac{9}{2} \frac{(\alpha - \sin \alpha)^2}{(1 - \cos \alpha)^3}~.
    \label{tophatdens}
\end{equation}
The parameter $\alpha$ is related to redshift via,
\begin{equation}
    \frac{1 + z_{\rm col}}{1 + z} = \left(\frac{\alpha - \sin \alpha}{2 \pi}\right)^{2/3}~,
    \label{tophatredsh}
\end{equation}
where $z_{\rm col}$ (corresponding to $\alpha = 2 \pi$) is the redshift at which the halo would collapse to a singularity (in reality non-spherical distributions, shell crossings, and shock waves would prevent this extremal collapse). The halo is considered to reach a state of virialization when the halo radius is half of its maximum, with the overdensity $1 + \delta_{\rm vir} = 18 \pi^2$ at the redshift $z_{\rm col}$. Comparing this virialization radius with the top-hat radius we find,
\begin{equation}
    r = \frac{r_{\max}}{2} (1 - \cos \alpha)~,
\end{equation}
so that virialization is reached when $\alpha = 3\pi/2$, or in terms of redshift, 
\begin{equation}
    z_{3\pi/2} = 1.06555(1 + z_{\rm col}) - 1~.
\end{equation}
We assume that a baryon overdensity rapidly forms around the dark matter overdensity after decoupling. \equasref{tophatdens}{tophatredsh} can be used to solve this overdensity numerically for $z_{\rm decouple} > z > z_{3\pi/2}$.

\subsection{Beyond virialization}
\noindent  The spherical top-hat model is no longer valid for the collapse of the cloud beyond virialization. There are numerous key processes, such as the formation of molecular hydrogen, to which the collapse of the cloud is sensitive and we must carefully model the baryon and dark matter densities beyond this point. Notably, the dynamics of the baryons decouples from the dark matter---the collapse of the baryonic halo during the period is driven by dissipative cooling, whereas the profile of dissipationless dark matter changes only due to the varying gravitational potential of baryons.

In \refsref{Birnboim:2003xa}{Dekel:2004un}, a simple analytic model is introduced to calculate the baryon profile. This model assumes no baryon shell crossing and short cooling timescales so that the baryons essentially free fall into the center, while the dark matter distribution is taken to be a constant isothermal profile. The baryon velocity can be computed using conservation of energy:
\begin{equation}
    \frac{1}{2} \left(\frac{d r}{d t}\right)^2+\phi(r)=\frac{1}{2} v_{\mathrm{vir}}^2+\phi\left(r_{\mathrm{vir}}\right)~.
    \label{req}
\end{equation}
Here we adopt the approach in the one-zone model \cite{Omukai:2000ic} by treating the baryon density to be uniform, so that we are only required to track the radius of the outermost mass shell. The gravitational potential at this shell is given by,
\begin{equation}
    \phi(r) = -\frac{G M_h}{r_{\rm vir}}\left[f_b \frac{r_{\rm vir}}{r} + (1 - f_b)(1 + \ln \frac{r_{\rm vir}}{r})\right]~,
\end{equation}
where $M_h$ is the halo mass (including both baryons and dark matter) and $f_b = \Omega_b / \Omega_m$ is the baryon mass fraction. The baryon density can be straightforwardly calculated once we solve the differential equation for the radius $r$ in~\equaref{req}.

As the mass shell contracts, the heating rate is enhanced as the baryons sink into the core of the dark matter halo and attenuate more of the radiation supplied from the heat source. In addition, the dark matter profile responds to the infalling baryonic component with the contraction of its orbits, increasing its density (and therefore increasing the heating rate of the cloud). It is then crucial to take into account the subsequent contraction of the dark matter halo to obtain the PBH heating rate, and the constant isothermal dark matter profile will not be sufficient for this purpose.

\subsection{Dark matter response to collapse}
\noindent The response of the dark matter to the baryonic in-fall can be modeled by adiabatic contraction~\cite{Steigman:1978wqb, Zeldovich:1980st, Ryden:1987ska}. The key insight of adiabatic contraction is that the mass enclosed by a shell of radius $r$ has an adiabatic invariant $M(r) r$ in a slowly varying potential. This allows us to relate the contracted dark matter profile $M_{DM}(r)$ to the initial profile $M_i(r_i)$ (including baryons and dark matter) via the relation,
\begin{equation}
    r (M_b(r) + M_{DM}(r)) = r_i M_i(r_i)~.
    \label{ac}
\end{equation}
This equation can be solved given a final baryon profile and this so-called `steady state' dark matter density solution typically has a power law dependence on the baryon density outside the core region~\cite{Spolyar:2007qv}. Motivated by this result, we adopt a power law dependence during the entire collapse period:
\begin{equation}
    \rho_{DM}(z) = \rho_{DM}(z_{\rm vir}) \left(\frac{n_b(z)}{n_b(z_{\rm vir})}\right)^{\zeta}~,
    \label{rdmst}
\end{equation}
We take $\zeta$ to be a free parameter here, noting that  Ref.~\cite{Spolyar:2007qv} found $\zeta \sim 0.81$. This time-dependent dark matter density is not fully consistent with the baryon density model, since $\phi(r)$ is calculated here in a constant dark matter background. Nevertheless, we can still use \equasref{req}{rdmst} as benchmark values---indeed there is a very good agreement between the estimated baryon density and numerical simulations~\cite{Birnboim:2003xa}.

\begin{figure}[!ht]
    \centering
    \captionsetup{width=.95\columnwidth}
    \includegraphics[width=.95\columnwidth]{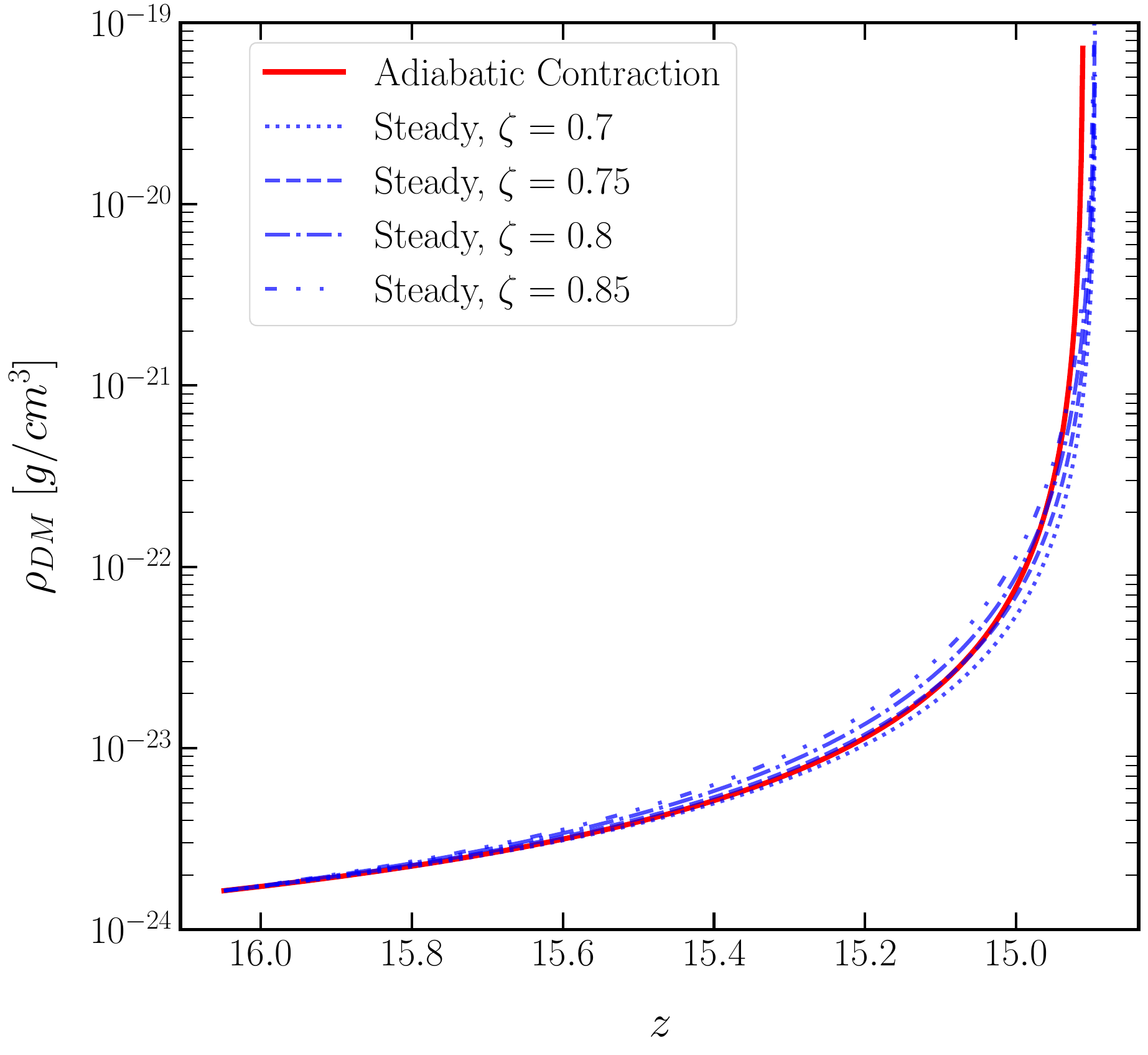}
    \caption{Evolution of the dark matter density during the halo collapse at $z\sim 15$ in both the `adiabatic contraction' (red) and `steady-state' (blue) approximations. The initial halo radius is taken to be $100~pc$ at $z = 1100$.}
    \label{fig:DM_dens}
\end{figure}

A slightly more accurate approach would be to treat ~\equaref{ac} dynamically during collapse, instead of only solving it for the final steady state. We still take the initial profile at virialization to be isothermal, so that the adiabatic contraction now requires,
\begin{equation}
    r f_b M_h+ r\left(1-f_b\right) M_h \frac{r_i}{r_{\rm vir}}=M_h \frac{r_i^2}{r_{\rm vir}}~.
\end{equation}
This equation can be used to solve for $r_i$, the initial radius of the dark matter shell that is contracted to the current baryon halo radius $r$. The dark matter density inside the baryon halo can be crudely approximated as
\begin{equation}
    \rho_{DM}(z) = (1 - f_b) M_h \frac{3 r_i}{4 \pi r_{\rm vir}r(z)^3}~,
\end{equation}
and the baryon shell evolves as,
\begin{equation}
    \frac{d^2 r}{d t^2} = - \frac{G(f_b M_h + (1-f_b)M_h r_i /r_{\rm vir}) }{r^2}~.
\end{equation}
We refer to this method as the `adiabatic contraction' approximation (in contrast to the `steady-state' method). We plot numerical solutions to the two different approaches in Fig~\ref{fig:DM_dens} for different values of the power~$\zeta$. We find very good agreement between these models during the important initial stages of the collapse. The slight mismatch at high densities is expected---the contracted dark matter halo will source a deeper potential and lead to an increased baryonic infall, which in turns causes further dark matter contraction, so both the dark matter and baryon densities should ultimately be higher compared to the steady state model. 

Finally, we point out that for each of the two approximation methods outlined here, we do not keep track of the portion of the dark matter halo that is outside the baryonic shells---all of the dark matter density used in this work refers to the density within the baryon halo. This gives us a conservative estimation of the total heating rate, since there may actually be a substantial population of dark matter exterior to the baryonic cloud which contributes to its overall heating.

\subsection{Temperature evolution}

\noindent The temperature of the halo in the presence of additional heating sources evolves as,
\begin{equation}
    \frac{dT}{dt} = (\gamma - 1)\left( \frac{\Dot{n}_b}{n_b} T + \frac{L_{\rm cool} + L_{\rm heat}}{k n_b} \right)~.
    \label{temp}
\end{equation}
The first term on the right hand side is commonly referred to as the `adiabatic heating/cooling' term since it corresponds to the temperature change in an adiabatic compression/expansion process. The function $L_{\rm cool}$ includes the additional cooling from the gas components, while $L_{\rm heat}$ is the additional heat source that we will discuss in detail in Sec.~\ref{sec:pbh}. In principle, the adiabatic index $\gamma$ is not a constant as the chemical components of the gas change over time (in particular the density of the diatomic component $\mathrm{H}_2$ changes). However, as the dominate component is always neutral hydrogen atoms, we can neglect this effect and we refer to \refref{Stasielak:2006br} for a detailed derivation of this equation including a time dependent~$\gamma$.

We should note that in the literature it is common to use $n_b$ for both the hydrogen number density and the baryon number density, while the latter also contains free electrons and helium. Here, we make the distinction by denoting the hydrogen density as $n$.

Before the first generation of stars is born, the primordial gas cloud is metal free. The chemical components of the gas include atomic (H) and molecular hydrogen ($\htwo$), ionized hydrogen ($\rm H^+$) and free electrons. Helium and trace amounts of lithium produced from BBN do not play important roles for the gas cooling. The cooling function generally depends on the density of the contributing components. The total hydrogen density can be decomposed into these components,
\begin{equation}
    n = n_H + n_{H^+} + 2 n_{H_2}~,
\end{equation}
while the fraction of the $i$-th component is simply, 
\begin{equation}
    x_i = \frac{n_i}{n}~.
\end{equation}

The cooling channels we consider fall into the following categories, where the rates are all in units of $\rm J\ cm^{-3}\ s^{-1}$:

1) Adiabatic cooling (heating): This channel is driven by the initial adiabatic expansion of the gas, as reflected in the first term on the right hand side of \equaref{temp}. During the collapse phase, the gas is heated instead.

2) Inverse Compton cooling: The thermal electrons in the gas are scattering by CMB photons and lose energy, with the rate given by \cite{Haiman:1996wue}:
\begin{equation}
    L_{ic} (T) = 1.017\times 10^{-44}  \Tcmb^4 (T - \Tcmb) x_e n~,
\end{equation}
in terms of the CMB temperature $\Tcmb$.

3) Hydrogen line cooling: Atomic hydrogen can cool via collisions with free electrons. We use the cooling rate in \refref{Sethi:2008eq}:
\begin{align}
    L_{HI}(T) = &7.9\times10^{-26} \left(1 + \left(\frac{T}{10^5\ \rm K}\right)^{0.5}\right)^{-1} \nonumber\\
    \times&\exp{(-118348\ \rm K/T)} n_e n_{H}~.
\end{align}

4) Molecular hydrogen cooling: Similar to atomic hydrogen, molecular hydrogen can also cool via collisional excitations. Due to its fine spaced rotational-vibrational energy levels, $\htwo$ can cool effectively at low temperature while hydrogen line cooling is exponentially suppressed below $\sim10^4$ K. The cooling rate is taken from \refsref{Galli:1998dh}{hollenbach1979molecule}. This is the most important cooling channel to suppress in order to prevent the fragmentation of the cloud as it collapses.

We only include the dominant cooling channels for each components in the temperature range of interest. For a more complete compilation of cooling channels, see \refref{Stasielak:2006br}. We plot the cooling channels together with PBH heating in~\figref{fig:results1a} in the case of successful direct collapse (high heating) and failed direct collapse (low heating). The major difference resides in the hydrogen line and molecular hydrogen cooling. In the high heating case, the formation of $\htwo$ is suppressed and hydrogen line cooling dominates almost the entire time and maintains the halo temperature at around $10^4~K$, as we will see explicitly in the following sections. In the failed case, molecular hydrogen cooling takes over, counteracting the PBH heating throughout the majority of the collapse process.

\afterpage{
\begin{figure}[!ht]
    \centering
    \begin{subfigure}[b]{\textwidth}
        \includegraphics[width=.7\paperwidth]{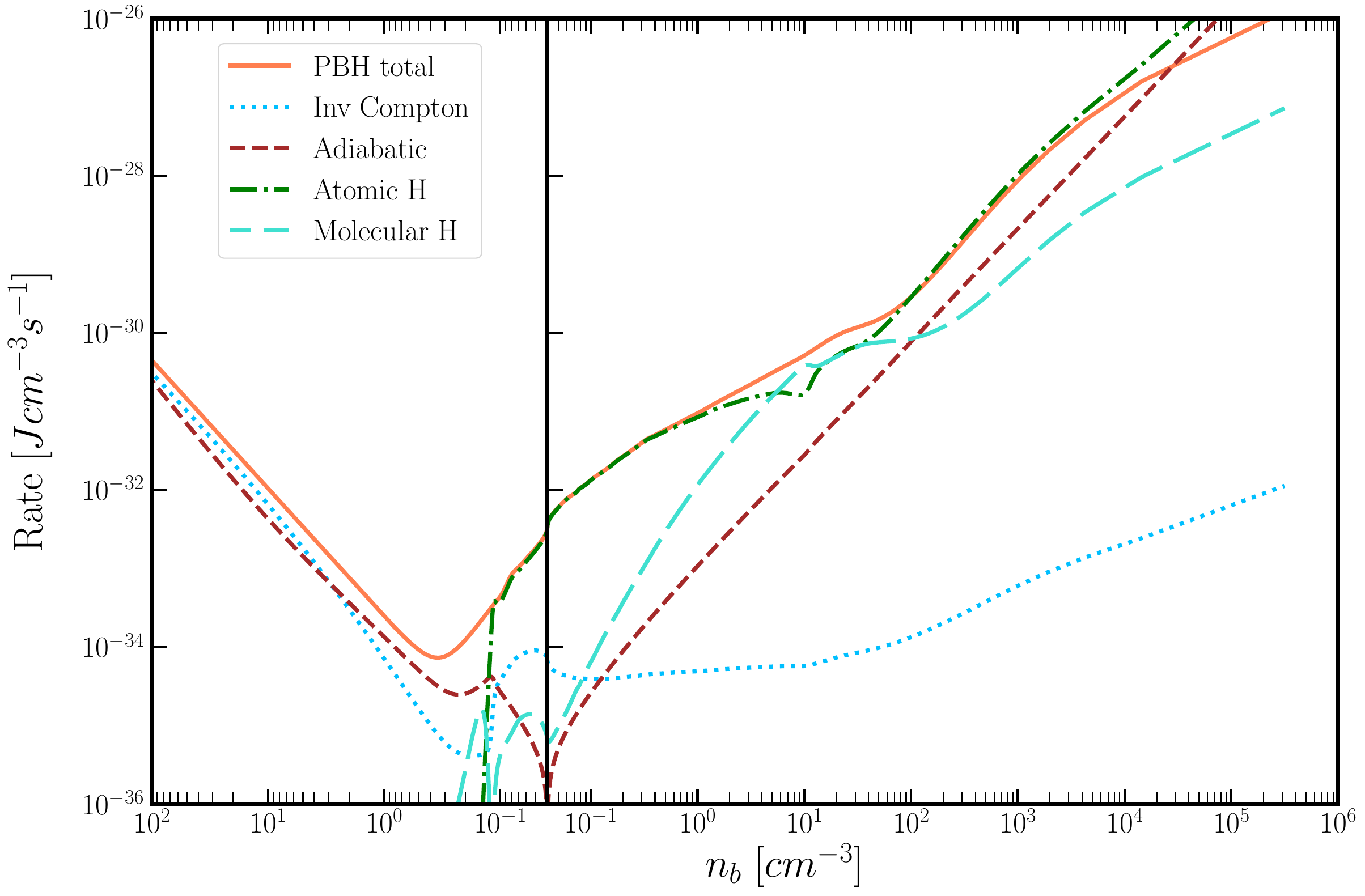}
    \end{subfigure}
    \begin{subfigure}[b]{\textwidth}
        \includegraphics[width=.7\paperwidth]{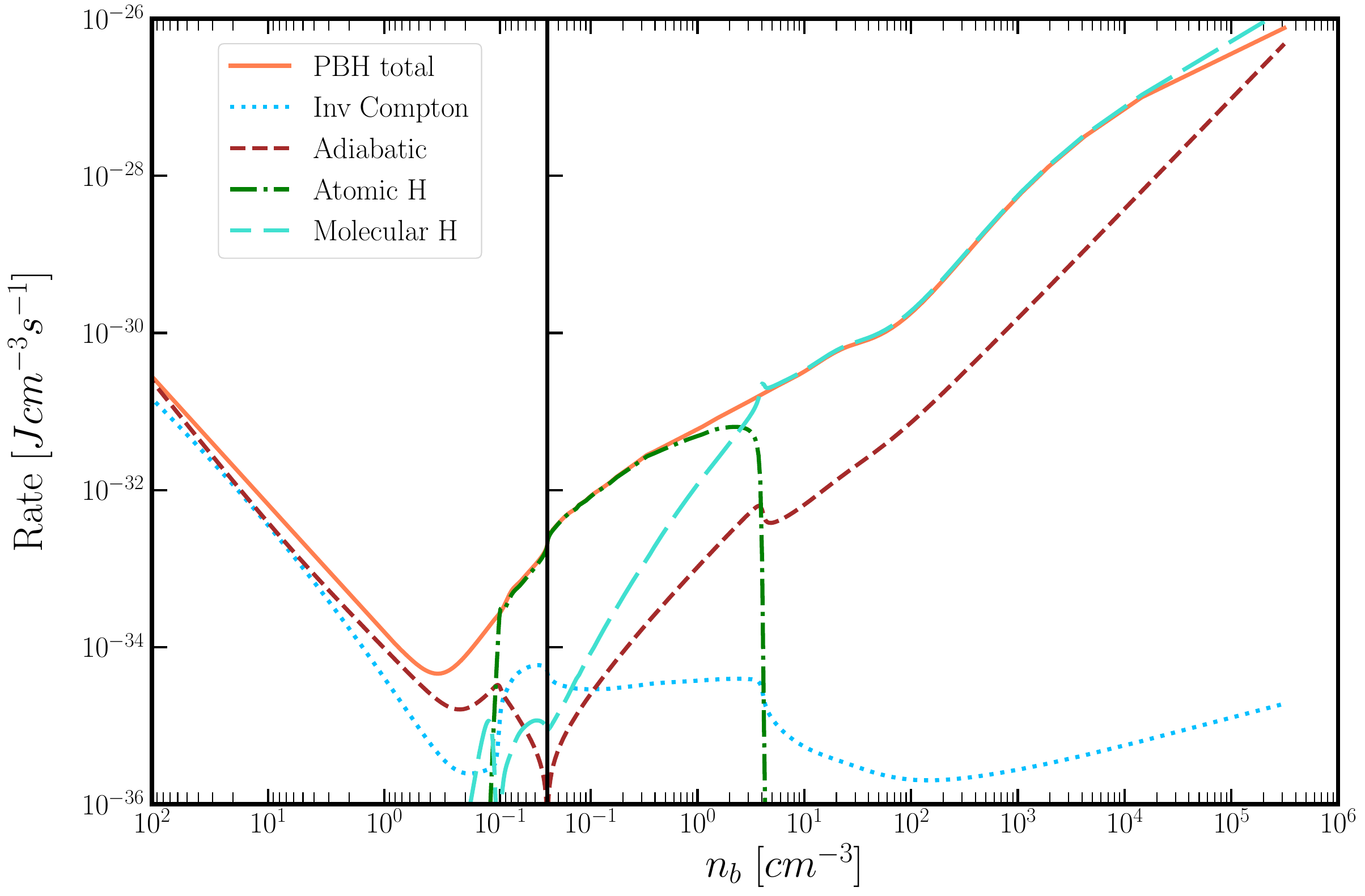}
    \end{subfigure}
    \captionsetup{width=\textwidth}
    \caption{Various sources of heating and cooling for the collapsing cloud as a function of baryon density in the case of sufficiently high (above) and too low (below) PBH densities  such that direct collapse can occur. In the sufficiently high case, we have set $f_{\rm BH, in} = 1.6 \times 10^{-4}$, whereas $f_{\rm BH, in} = 1.0 \times 10^{-4}$ for the too low case.  The `PBH total' curve describes the heating of the cloud due to the PBH secondary Hawking radiation. We can see that the molecular hydrogen cooling channel comes close to overtaking the PBH heating, but just misses, allowing the collapse to proceed without fragmentation. The time direction of collapse is read from left to right, spanning from $z\sim1100$ until the rapid collapse at $z\sim 20$ in this case. The PBH lognormal mass function is taken to be centered at $2.0 \times 10^{14}~g$ with $\sigma = 0.2$.  }
    \label{fig:results1a}
\end{figure}
\clearpage
}

\subsection{Chemical evolution}

\noindent Finally, we must carefully track the heavily coupled chemical evolution of the cloud. As we have already emphasized, reducing the density of $\htwo$ is an essential benchmark for successful direct collapse to a black hole. In this work, we keep track of the hydrogen in atomic or molecular form, ionized hydrogen (protons), free electrons and helium. For simplicity, we neglect the effect of He or $\htwo$ ionization, so that $n_e = n_{H^+}$. 

BBN predicts the helium mass fraction to be $Y = 0.244$, giving the baryon number density in terms of the hydrogen density:
\begin{equation}
    n_b = n\left(1 + \frac{Y}{4X} - x_{H_2} + x_e \right)~.
\end{equation}
The electron number density is affected by the photo- and collisional-ionization of hydrogen atoms. For the following, we use the reaction rates $k_i$ compiled in \refref{2010MNRAS.402.1249S}, unless stated otherwise. The free electron fraction evolves as,
\begin{equation}
    \frac{d x_e}{dt} = k_{22} x_H - k_4 x_e^2 n + k_1 x_e n_H~.
    \label{xe}
\end{equation}
In this equation, we did not include the Peebles factor $C$ commonly used in recombination calculations~\cite{Peebles:1968ja}. This is valid as long as we work at lower redshifts ($z < 1000$) as $C \sim 1$ in this regime. The additional term with $k_1$ comes from the collisional ionization of H with $e^-$. 

We note here that when solving the chemical equations, we did not consider the additional effects from the direct injection of particle species into the cloud from the heating source. Rather, the particles injected only enter the halo evolution via the energy injected into the cloud as heat. The impact of this influx of photons, electrons, and protons would certainly impact the chemical evolution nontrivially, although we expect the effect on quantities like the ionization fraction to be overall subdominant, since the PBHs themselves are still a subdominant component of the cloud by mass. We discuss the difficulties of evolving the full chemical system with the primary Hawking radiation products in the following section.

Next, we consider the formation of $\htwo$ in the cloud. The dominant channel is a two step process:
\begin{align}
    H + e^- &\rightarrow H^- + \gamma, \\
    H^- + H &\rightarrow H_2 + e^-~.
\end{align}
Other formation channels of $\htwo$ involve the intermediate products $\rm H_2^+$ or $\rm He H^+$. \refref{Hirata:2006bt} showed that the $\rm H_2^+$ and $\rm He H^+$ channels contribute to less than two percent of the final $\htwo$ abundance, so we only included the $\rm H^-$ channel in our computation. The $\rm H^-$ abundance is governed by the following equation:
\begin{align}
    \frac{d x_{H^-}}{dt} = &k_{9} x_H x_e n - k_{10} x_H x_{H^-} n - k_{13} x_e x_{H^-} n \nonumber\\
    -&k_{19} x_e x_{H^-} n - k_{20} x_H x_{H^-} n - k_{\gamma} x_{H^-}~,
\end{align}
where all the rates follow the conventions of~\refref{2010MNRAS.402.1249S}, with the exception of the photodissociation rate $k_{\gamma}$ for the process $H^- + \gamma \rightarrow H + e^-$. Using the principle of detailed balance, we can relate $k_{\gamma}$ to its inverse process by,
\begin{align}
    k_{\gamma}(T_{\rm cmb}) = &4 \left(\frac{m_e k_b \Tcmb }{2\pi \hbar^2}\right)^{3/2}\nonumber\\
    \times&\exp{(-0.754 \mathrm{eV} / (k_b \Tcmb))} k_{9}(\Tcmb)~.
\end{align}
The equilibrium fraction of $\rm H^-$ is given by,
\begin{equation}
    x_{H^-} = \frac{k_{9} x_H x_e n}{k_{\gamma} +(k_{13} + k_{19}) x_e n + (k_{10} + k_{20}) x_H n}~.
\end{equation}
Then the $\htwo$ abundance can be computed using the equilibrium $\rm H^-$ fraction, which evolves as,
\begin{equation}
    \frac{d x_{H_2}}{dt} = k_{10} x_H x_{H^-} n - k_{15} x_H x_{H_2} n - k_{18} x_e x_{H_2} n~.
    \label{xh2}
\end{equation}
Here we include the destruction of $\htwo$ by collisional dissociation with $H$ and $e^-$. For the rate $k_{15}$, we use the simpler functional forms of \refsref{shapiro1987hydrogen} {Palla:1983nv}. Collisional dissociation with $\htwo$ is neglected because of the relatively low abundance of $\htwo$. 

We solve \equassref{temp}{xe}{xh2} numerically with the density evolution tracked by the procedure described in previous sections. It is convenient to work with redshift $z$ as our time variable, and its relation with time is essentially given by the standard cosmological relation,
\begin{equation}
    \frac{dt}{dz} = - \frac{1}{H_0 (1+z)\sqrt{\Omega_{\Lambda} + (\Omega_{dm} + \Omega_{b})(1+z)^3}}~.
\end{equation} 
 We start the evolution at the end of the decoupling where $z = 1100$, $x_e = 0.01$, and with initial $\htwo$ fraction set to zero.

\section{PBH Heating}\label{sec:pbh}
\noindent We consider now explicitly the heating of the collapsing cloud by the evaporation of a relatively sparse population of small, exploding primordial black holes. For simplicity, we model our PBH population with a lognormal distribution given by,
\begin{align}
    \frac{\mathrm{d}n(M)}{\mathrm{d}M} = \frac{n_\mathrm{BH}}{\sqrt{2\pi}\sigma M} \exp{\left( -\frac{(\ln (M/M_*))^2}{2\sigma^2} \right)}~,
\end{align}
which is a reasonable fit for realistic PBH formation processes, particularly the collapse of density perturbations from inflation~\cite{Green:2016xgy,Kannike:2017bxn,Calcino:2018mwh,DeRocco:2019fjq,Carr:2020gox}. For the lognormal distribution, the total black hole number density $n_\mathrm{BH}$ is related to the fraction of dark matter in PBHs $f_\mathrm{BH}$ by,
\begin{align}
    n_\mathrm{BH} = \frac{\fbh \rho_\mathrm{DM}}{M_*\exp\left(\sigma^2 /2\right)}~.
\end{align}
Since we are interested in evaporating black holes, we are also careful to correctly evolve this spectrum in time as the lightest black holes in the distribution explode earliest~\cite{Mosbech:2022lfg}.

Constraints exist for extended PBH mass functions down to about $10^{15}~g$, where the lifetime is close to the age of the universe~\cite{Carr:2020gox,Carr:2017jsz}. Primarily these constraints come from (non-) observation of the Hawking spectra, particularly gamma rays and electrons~\cite{Korwar:2023kpy,Boudaud:2018hqb}. For black holes smaller than this, which evaporate before today, observations must look to the earlier universe. Some of these constraints have been compiled in the monochromatic case in Fig.~11 of Ref.~\cite{Carr:2020gox}, although they miss the important Ref.~\cite{Cang:2021owu} for the relevant masses in this paper. The primary source of these constraints come from CMB anisotropies, spectral distortions, and the 21-cm line~\cite{Acharya:2020jbv,Chluba:2020oip,Cang:2021owu}. 

For smaller black hole masses, not all of the early universe constraints have been computed for extended mass functions. Generally, however, monochromatic constraints are neither significantly avoided nor made significantly stronger when converting to well-behaved (and not-too-wide) extended distributions. For simplicity then, we will make choices for $\fbh$ that should be \textit{approximately} allowed by existing observations, at least for not-too-large widths $\sigma$. As we will see, we are probably making very conservative estimates of the PBH heating, for a number of reasons, so our approximation of the allowed $\fbh$ is presumably not problematic. 

For our purposes we consider lognormal mass functions with relatively narrow ($\sigma\leq0.5$) widths, with $\fbh\lesssim 10^{-10}$ and centered at the black hole mass $M\sim10^{14}~g$, so that the PBH population is only a significantly subdominant component of the dark matter (before it evaporates completely). This central mass is chosen so that the black holes are predominantly exploding between the redshifts $z=1000$ and $z=10$, and we also note that black holes of these sizes are far too small for accretion to be a relevant process.

\subsection{Local PBH abundance}
\noindent In order to source the necessary heating to prevent fragmentation, we must entertain the possibility that the cloud could have a locally larger fraction of dark matter in PBHs than the background, so that we have both an `internal' $\fint$ and `external' $\fext$. The additional clustering of PBHs around denser regions can be motivated in a number of ways, and we note that we are not subject to any isocurvature constraints~\cite{Young:2015kda,Tada:2015noa,Young:2019gfc} on PBH dark matter in this case, since the PBHs in total are a significantly subdominant population of the dark matter.

The `classic' PBH formation mechanism---where black holes form from overdensities after inflation---is extremely sensitive to slight variations in the form of these perturbations. (indeed, the physics of this collapse is far from settled today~\cite{Franciolini:2023wun}). In particular, under the assumption that the initial perturbations are Gaussian-distributed, there is no PBH clustering besides Poisson fluctuations. However, as noted in Refs.~\cite{Ferrante:2022mui,Franciolini:2023wun} among others, virtually all curvature perturbations which lead to PBH formation are non-Gaussian, especially in the small-scale regime where PBHs form (as opposed to the larger scales of the CMB fluctuations). Not only do these non-Gaussianities affect broadly the PBH constraints and mass distributions~\cite{Ferrante:2022mui}, but they also lead to enhanced clustering of the black holes~\cite{Young:2019gfc,Carr:2023tpt}. More simply put, PBH are more likely to be formed on top of larger superhorizon density contrasts (when these are sourced by non-Guassianities). Indeed, in Fig. 2 of Ref.~\cite{Young:2019gfc} it is demonstrated that for relatively small values of the product of the Gaussianity parameter and the superhorizon curvature perturbation, there can be local increases in the PBH fraction of many orders of magnitude. In particular, for a curvature perturbation $\zeta$ and non-Gaussianity parameter $f_\mathrm{NL}$ such that $\zeta f_\mathrm{NL}\sim 0.5$, it is possible to get a local PBH fraction $\sim10^7$ times larger than the background.

Secondly, PBHs may form by mechanisms other than the collapse of perturbations after inflation. In particular, if the black holes form out of some dynamics in the dark sector, then they would naturally be found more readily in the dark matter halos which are associated to the baryonic clouds in question. This could be the case, for example, if the PBHs formed from Yukawa interactions in the dark sector~\cite{Flores:2020drq,chakraborty_formation_2022,Lu:2022jnp,kawana_primordial_2022,Domenech:2023afs,Flores:2023zpf}, or if the dark matter is primarily composed of extended objects such as Q-Balls or oscillons~\cite{Coleman:1985ki,Kusenko:1997si,Bogolyubsky:1976nx,Gleiser:1993pt} whose interactions could lead to PBH formation~\cite{Cotner:2016cvr,Flores:2021jas,Cotner:2018vug}, or if there is some dark phase transition which pushes dark matter into PBHs~\cite{Baker:2021nyl} (analogously to quark nugget formation~\cite{Witten:1984rs}). Indeed, in some of these scenarios it is possible to form PBHs of any size at much later times~\cite{Picker:2023lup,Picker:2023ybp} so that we could actually choose a smaller-mass population of PBHs to source the heating of the cloud---which could be an interesting followup to this work. All of the above represents just a few possibilities for significant PBH clustering, but it is not difficult to justify that black holes---which form from small-scale overdensities---might be more likely to form in regions of higher dark matter density.

Of course, we must also be careful to not populate \textit{too} many dark matter overdensities with large PBH fractions compared to the background, or else the total PBH fraction $\fbh$ would become too large and we would need to carefully reevaluate existing constraints for this highly-clustered scenario. Indeed, we anyway presumably do not want every molecular cloud to form a SMBH, although the precise proportion of clouds that should collapse to black hole will need to be determined with continuing high-redshift observations. For the purposes of this paper, we presume that we are considering either the most massive clouds, or the ones that just happen to inhabit the `high-$\fint$ tail' (or both). Since we neither have precise theoretical calculations for the number of halos which might have large $\fint$, and we don't know precisely the rate of high-redshift SMBH formation, it remains challenging to incorporate our model into SMBH density calculations such as \refref{Buchner:2019hma}, which usually take certain simplifying assumptions. Therefore, we leave the estimation of this abundance of this population to future works. Instead, we demonstrate merely that it is \textit{possible} for clouds to directly collapse to black holes with the assistance of exploding PBHs. Such a calculation could even be used as a constraint on the PBH population once we have a better understanding of the required SMBH collapse rate (and predictions for the PBH clustering).

\subsection{Heating with Hawking radiation}

\noindent Black holes radiate a thermal spectrum consisting of all particles with a mass below their temperature. The particle emission rate is given by,
\begin{equation}
    \frac{d^2 N_{i}}{d t d E} =  \frac{1}{2\pi}\sum_\mathrm{dof}\frac{\Gamma_{i}(E,M,a^*) }{e^{E'/T} \pm 1}~,
\end{equation}
where $N_i$ is the number of particles emitted, $\Gamma_i$ is the so-called `greybody factor', $E'$ is the energy of the particle, $a^*$ is the reduced spin parameter, the sum is over the degrees of freedom of the particle, and the $\pm$ sign refers to fermions and bosons respectively. We use the code BlackHawk with its associated particle physics packages~\cite{arbey_blackhawk_2019,arbey_physics_2021,Sjostrand:2014zea,Bellm:2015jjp,Coogan:2019qpu,Bauer:2020jay} to compute these spectra.

We estimate the heating of the cloud from these evaporating black holes in the following way. In a vacuum, most of the primary Hawking particles decay or annihilate with themselves, producing a flux of secondary particles consisting of photons, electrons, neutrinos, and protons. Following Freese et al.'s treatment of dark star energy injection~\cite{Freese:2015mta}, we then estimate the attenuation of these secondary particles in the cloud (since some of the Hawking radiation passes right through the cloud in this lower density regime). There are two components heating the cloud---the global background of evaporating PBHs which uniformly heats the universe, and the heating from the `internal' population of black holes which are part of the collapsing cloud. Which of these dominates the heating depends on their relative densities and the density of the baryonic matter in the cloud, which generally attenuates more of the Hawking radiation as it becomes more dense.

We estimate the attenuation fraction for photons passing through the cloud as,
\begin{align}
    f_{Q,\gamma} &= 1- \exp\left(-X/X_0\right)\nonumber\\
    X &\equiv 1.2~m_p~n_b~r~,
\end{align}
where $X_0\sim 100~g~cm^{-2}$~\cite{ParticleDataGroup:2020ssz,ParticleDataGroup:2004fcd} at the relevant photon energies here and $m_p$ is the proton mass, $r$ is the cloud radius and $1.2$ is the average distance from a point inside the unit sphere to its surface~\cite{stackexchange} (for the exterior PBH population we would better use the average sphere chord length $4/3$~\cite{dirac_lol,SJOSTRAND20021607} but this will make little difference to our estimation). We estimate that the photons which are attenuated by the cloud transfer their entire energy into it, heating it up. We also reiterate our earlier point that a complete analysis would additionally compute the contribution of the photons to the full chemical system of the cloud.

For the electrons, there are two regimes, following the estimations of Ref.~\cite{Freese:2015mta}. Lower energy electrons ($E\lesssim 280~MeV$) lose energy by ionization, so we estimate their attenuation fraction $f_{Q,e\downarrow}$ exactly as for the photons but with $X_0 = E/(4.4\times10^{-3}~GeV)~g~cm^{-2}$, where $E$ is the electron energy. Higher energy electrons produce an electromagnetic cascade, where the attenuation fraction is given by,
\begin{align}
    f_{Q,e\uparrow} &= \gamma\left(a,X/2X_0\right)/\Gamma\left(a\right)~\nonumber\\
    a &= 1+\frac12\left(\ln\left(E/280~MeV\right)-1/2\right)~,
\end{align}
where $\gamma(x,y)$ is the incomplete gamma function, $X_0=63~g~cm^{-2}$ in hydrogen, and $X$ is defined as before.

We assume that the neutrinos have no effect on the heating. We also assume that the majority of secondary protons are non-relativistic in this scenario and so are fully attenuated by the clouds over the lengths relevant for our analysis, adding additional heating to the clouds. If the black holes were significantly smaller than the $\sim10^{14}~g$ used here, the protons would however be relativistic and their attenuation would need to be explicitly computed.

\subsection{Primary vs. secondary Hawking spectra}

\noindent Using just the secondary Hawking spectra as the heating source is a crude approximation to the dynamics of the exploding PBHs inside the cloud. To be fully consistent we would need to take the \textit{primary} Hawking spectra and calculate its interaction with the full cloud, including the energy injected from the collisions of primary Hawking particles and hadronic jets with the molecules in the cloud and the subsequent (and coupled) effects on the cloud's chemical system. This is a very complicated scenario, both in terms of the physics which must be included, and the required dependency on the current state of the molecular cloud, adding enormous computational expense to the numerical simulations.

In light of this, we opted to merely check if we could achieve the conditions for direct SMBH formation with just the secondary radiation. The fully-worked scenario is likely to feature a somewhat larger deposition of energy into the cloud, since the annihilation of primary particles with the cloud would directly heat it---our estimation, meanwhile, first converts these primary particles into stable secondary particles, which only deposit \textit{some} of their energy via the fraction that is attenuated within the cloud. We then conclude that our computation here is again a conservative estimation of the PBH heating rate.
\begin{figure*}[!ht]
    \centering
    \includegraphics[width=.8\paperwidth]{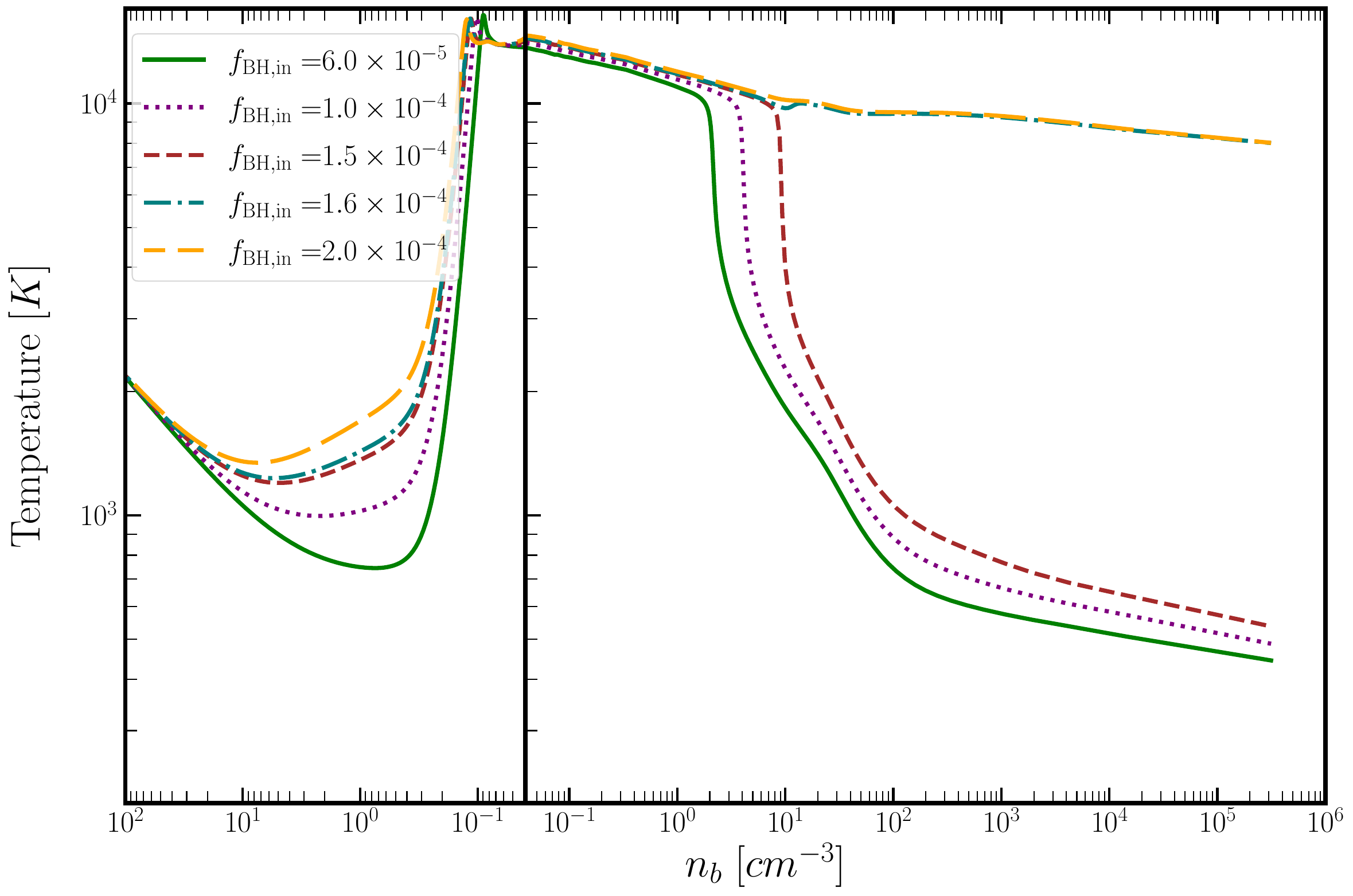}
    \caption{Cloud temperature as a function of baryon density, with different lines representing different values of the internal PBH fraction $\fint$. In order for direct collapse to proceed, the temperature needs to stay roughly around $10,000~K$. We can see how sensitive the collapse is to this condition---for even slightly lower heating, the emergence of the $\htwo$ cooling channel rapidly cools the cloud and prevents SMBH formation. Here the PBH parameters are $M_* = 2.0\times 10^{14}~g$ and $\sigma = 0.2$.}
    \label{fig:results2}
\end{figure*}

\begin{figure*}[!ht]
    \centering
    \includegraphics[width=.8\paperwidth]{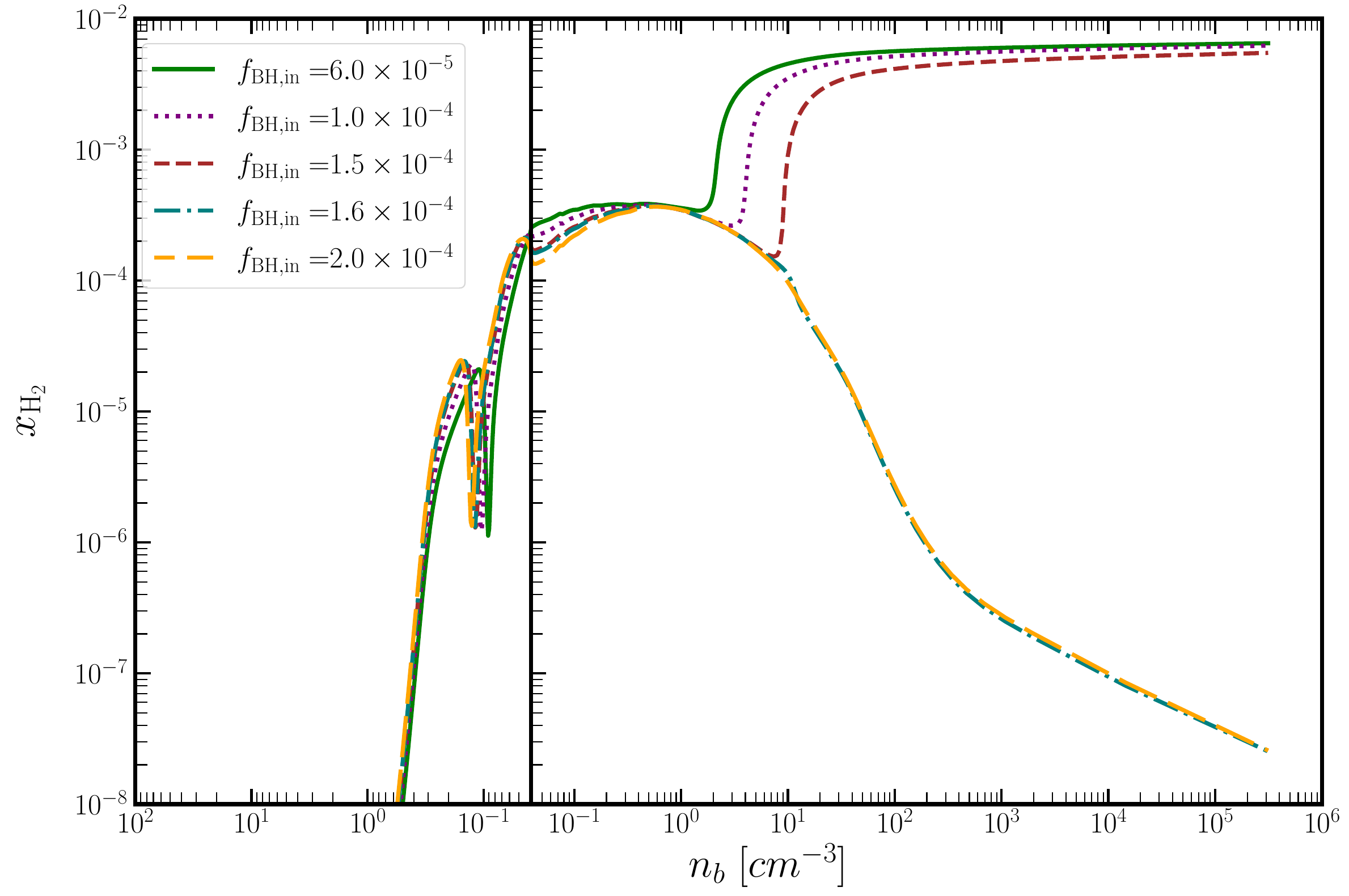}
    \caption{Fraction of $\htwo$ as a function of baryon density during the collapse for a handful of values of internal PBH fraction $\fint$. In the two cases where the heating is sufficient for direct SMBH collapse we can see the fraction of molecular hydrogen plummet. Here the PBH parameters are $M_* = 2.0\times 10^{14}~g$ and $\sigma = 0.2$.}
    \label{fig:results3}
\end{figure*}

\subsection{Other potential heating sources}
\noindent The notion that dark matter (or a subcomponent thereof) could provide an additional heat source for these clouds is not entirely new---this has been studied before specifically in the context of annihilating weakly interacting massive particles dark matter~\cite{Freese:2015mta,Spolyar:2007qv}. In this scenario, as the collapsing cloud reaches very high densities, annihilating dark matter in the core of the cloud is able to source sufficient pressure to stop the gravitational collapse, forming speculative objects known as dark stars which could help explain some of the JWST observations~\cite{Ilie:2023zfv}. 

The dark star scenario relies on the density-squared dependence of the annihilation rate to source sufficient energy at high baryon densities, at what would otherwise be near the end point of the collapse. For this reason it would be very challenging to reproduce the dark star scenario with evaporating PBHs, where the energy injection scales linearly with their number density. Instead, we are interested primarily here in the relatively lower-density portion of the collapse, where Ref.~\cite{Omukai:2008wv,Inayoshi:2019fun} identified a `zone of no-return' where direct collapse of a gas cloud to a black hole was possible so long as the temperature remained sufficiently high at baryon densities of $n_b\sim 10^{4}~ cm^{-3}$. In this lower-density regime, it is easier to heat the cloud with a mechanism that scales with density, rather than density-squared, which only turns on much later.

We also could wonder if other dark matter candidates (or beyond the standard model physics) could serve in place of evaporating PBHs. For example, if there was a subdominant component of the dark matter which was decaying, this would inject energy in the form of decay products into the clouds. This is slightly harder to motivate than the PBH scenario for two reasons. Firstly, it is reasonable to expect the PBH population to form with an extended mass function, which would allow them to explode at a relatively uniform rate over a sufficiently long timespan (depending on the distribution width), whereas the population of a decaying particle decreases along an exponential curve. Secondly, and perhaps more importantly, our mechanism requires a locally higher fraction of the heating component inside the cloud than in the background. This is relatively easy to motivate with PBH clustering, but much more challenging with a decaying particle species.
\section{Discussion}\label{sec:results}

\noindent We numerically computed the evolution of a demonstrative halo to show the viability of our proposed mechanism. Specifically we take a halo set to collapse at $z\sim 20$ with mass $7.7\times10^8 \Ms$ (corresponding to an initial radius of $150~pc$ at $z=1100$), and with the demonstrative PBH parameters $M_* = 2.0\times 10^{14}~g$, $\sigma = 0.2$ and $\fext = 5.0\times 10^{-11}$. The central mass of the PBHs is chosen so that its lifetime is close to the redshift at collapse---the key period during which a high heating rate must be supplied to inhibit $\htwo$ formation. The exact value of $\fint$ which leads to the bifurcation behavior in Fig.~\ref{fig:results2} depends mildly on the central mass $M_*$ and $width$---since these control the energies of the particles as well as the epoch of PBH explosion---as well as the redshift of collapse (chosen as an initial parameter in the top-hat collapse model). For demonstrative purposes, we compute our results for one specific value of the black hole parameters and collapse model, but note that for other not-too-distant choices of parameters, we find similar behavior for the SMBH collapse with internal black hole fraction $\fint$ of the same order of magnitude. 

\subsection{Simulation results}

\noindent The results are shown in Figs.~\ref{fig:results2} and~\ref{fig:results3}---specifically, we plot the the cloud temperature and the $\htwo$ fraction in order to demonstrate the direct collapse to a black hole when the internal heating is sufficiently high. We observe a similar bifurcation behavior as in \refref{sethi2010supermassive}: the cloud is very sensitive to the internal heating rate and demonstrates drastically different behaviors even for a slight change of the heating rate near the critical $f_{\rm BH, in} \sim 1.6 \times 10^{-4}$ . Below this value, PBH heating is not sufficient to maintain the cloud temperature at $\sim 10^4~K$, so the molecular hydrogen cooling channel turns on and direct collapse to a SMBH is prevented. At the critical value, however, the halo collapses almost isothermally and the heating suppresses $x_{H_2}$ to a low value until reaching the critical density for the `zone of no-return'. Beyond this point, the halo continues to collapse to a supermassive star without fragmentation, regardless of the heating rates. General relativistic instabilities generically set in for supermassive stars with mass exceeding $\sim 10^5 \Ms$ \cite{Chandrasekhar:1964zz}, so that a supermassive black hole must be formed.

\subsection{On fine-tuning}
\noindent The success of direct collapse via this mechanism is sensitive to a number of parameters which are worth elucidating clearly. Arguably the most finely-tuned input is the acceptable PBH central mass range, since the black hole lifetime is quite sensitive to initial mass. Black holes evaporating with lifetime between $100$ and $400$ million years (roughly the range of interest, from redshifts $11-30$) have masses in the somewhat narrow range $1.4-2.2 \times 10^{14}~g$. However, because the black hole lifetime is so sensitive to the mass, even a relatively narrow mass function still results in a relatively broad timespan over which the PBHs are exploding. This allows the mechanism to work for the wide range of collapse times required to explain the JWST observations---this range of collapse times is embedded in the initial halo configuration, which we model as a top-hat collapse in which collapse time can be specified explicitly. 

This leads us to a second important point---the epoch of collapse is not determined by any PBH parameters. In our model, the exact collapse time is actually an input to the top-hat collapse model. Physically this can be understood as a statement that it is the properties of the halo itself (initial mass and size) which determine the time of collapse. The PBH heating is required to be significant enough to pass the threshold so that molecular hydrogen cooling never turns on, but otherwise does not impact the variety of SMBHs formed (in either formation time or mass function). Instead, other complicated but well-studied astrophysical properties related to the hosting halo, such as its mass, gas inflow, subsequent merger events, and accretion rates, are more significant to the final direct collapse black hole mass. In this regard, our scenario does not differ from the large body of existing literature on black hole direct collapse---we merely suggest a mechanism which suppresses molecular hydrogen cooling.

Finally let us return to the potential fine-tuning of the interior and total PBH fractions. The total PBH fraction is indeed subject to strong constraints, which is why we demand a high level of clustering in order to achieve sufficiently high internal fractions. We chose a demonstrative value of the total fraction which is unconstrained, but we could instead choose a smaller one and impose higher clustering---for example, in Ref.~\cite{Young:2019gfc}, there are many orders of magnitude to spare which would allow us to increase the clustering. As best as we know, there are no observational constraints on such high clustering for exploding PBHs on the relevant length-scales for this scenario---in the following section, we propose some potential future work on the detection of these exploding PBH 'hot spots'. Lacking such constraints we must conclude that there is no real fine-tuning needed for the clustering to be sufficiently high. Significantly, clustering is a statistical effect---some halos will have higher clustering and some less. The halos with sufficiently high clustering may then collapse to SMBHs. In principle, given a model of clustering, we could estimate from this the abundance of SMBHs---we leave such a calculation to future work.

\subsection{Other consequences of exploding PBH clusters}\label{sec:hotspots}
\noindent Before concluding let us briefly consider some potential additional consequences of clusters of PBHs which evaporate in the early universe. If the clusters were extremely large, they could leave an imprint on the CMB, or they could locally heat patches of the early universe. The phenomenology of such `hot-spots' has not been extensively studied but could feasibly affect BBN, the formation of dark matter, or even baryogensis, all of which have been studied in other PBH-related `hot spot' scenarios~\cite{Flores:2022oef,He:2022wwy,bellido2019RSPTA.37790091G}. 

One consequence of particular interest regards the molecular clouds with too little PBH clustering for direct collapse, but which still contain a reasonably large internal fraction $\fint$. Depending on the clustering statistics, this population might indeed be larger than the direct collapse population. In this case, there would be a number of early-universe molecular clouds with extra photon luminosity in the $\sim $MeV--GeV region from the Hawking evaporation of the PBHs, and after redshifting to today, this radiation would presumably be roughly within the X-ray range. It would certainly be interesting to search for these kind of anisotropies in the extragalactic X-ray background. 

Notably, the level of fluctuations in the cosmic infrared background  exceeds theoretical expectations~\cite{Kashlinsky:2002vf,Cooray:2012xj,Yue:2013hya,Matsumoto:2019stv,Kim:2019qli,Cheng:2022snx} on the angular scales $l\sim 10^3$.    An interesting possibility is that direct collapse to SMBHs can explain the fluctuation excess~\cite{Yue:2013hya}.  In our scenario, the fluctuations can receive an additional contribution from the small halos in which the PBH density is below the threshold for direct collapse, but in which the heating of gas by the MeV--GeV emission is still non-negligible.  The optical depth of a typical galactic bulge to MeV--GeV gamma radiation (which has the attenuation column depth $\lambda\sim 10-100$~g/cm$^2$) is $\sim 10^{-3}-10^{-2}$.  The heating of gas in halos with a high clustering of PBHs may result in additional contribution to infrared background fluctuations. 

\section{Conclusion}\label{sec:conc}
\noindent The origin of high-redshift black holes is a compelling mystery, dramatically brought into light by the recent observations by JWST of unexpectedly high-redshift quasars which challenge traditional explanations for the formation of SMBH in the early universe. We demonstrated here that a cloud of baryonic matter can be sufficiently heated by clustered exploding PBHs such that it collapses directly to a SMBH, even for PBH populations which are a subdominant portion of the dark matter. The amount of clustering of these PBHs depends sensitively on the exact details of their formation, but there is good reason to believe that in many of the most plausible scenarios for PBH formation, some portion of their population would indeed be highly clustered, leading to locally large densities within some collapsing gas clouds. It would be interesting in the future to determine the exact portion of the PBH population which is highly clustered and associated to a particular molecular cloud, so that we could estimate the SMBH formation rate by this mechanism. However, this is hampered both by theoretical uncertainties in the PBH formation mechanisms and the observational uncertainty in the true high-redshift SMBH abundance, and must be left to future work.

Our numerical work here is preliminary in a few ways, but the approximations we make throughout all err on the conservative side. As well as making various simplifications to the collapse process, we model the heating of the evaporating PBHs in a relatively naive way via the secondary PBH Hawking evaporation. A full simulation of the collapse, including the direct impact of the primary Hawking radiation in the full, coupled chemical system of the molecular cloud, would be required to make more detailed estimates of this scenario. Since this is both exceptionally complicated and computationally expensive, we opt here to show merely the viability of this mechanism, even with so many conservative approximations. In a more detailed computation, we expect to recover similar qualitative behavior to our numerical results here, but presumably for lower internal PBH fraction $\fint$.

\section*{Acknowledgements}
This paper was partially written on Tongva, Chumash, and Gadigal lands. 
We thank Katherine Freese, Philippa S. Cole and the anonymous referees for helpful discussions.
This work was supported by the U.S. Department of Energy (DOE) Grant No. DE-SC0009937. The work of A.K. was also supported by World Premier International Research Center Initiative (WPI), MEXT, Japan, and by Japan Society for the Promotion of Science (JSPS) KAKENHI Grant No. JP20H05853. 

This work made use of N\textsc{um}P\textsc{y}~\cite{numpy2020Natur.585..357H}, S\textsc{ci}P\textsc{y}~\cite{scipy2020NatMe..17..261V}, and M\textsc{atplotlib}~\cite{mpl4160265}.

\bibliographystyle{bibi}

\bibliography{main.bib}

\begin{thebibliography}{100}
\bibitem{stackexchange}
M.~Hardy, ``Average distance from a point in a ball to a point on its boundary.'' Mathematics Stack Exchange.
\end{thebibliography}

\end{document}